\documentclass[runningheads]{llncs}
\usepackage{graphicx}
\usepackage{amsmath}

\usepackage[labelfont=bf]{caption}                         % set caption font size
\usepackage[subrefformat=parens]{subfig}
\usepackage{enumitem}  
\usepackage{amssymb}
\usepackage[misc]{ifsym}
\usepackage{graphicx}
\usepackage{threeparttable}
\usepackage{ntheorem}
\usepackage{color}
\usepackage{colortbl}
\usepackage{multirow}
\usepackage{amsmath}
\usepackage{comment}
\usepackage{bm}                                            % bold math font
\usepackage{etoolbox}                                      % \newtoggle, \toggletrue, \iftoggle
\usepackage{url}
\usepackage{nth}
\usepackage{cite}
\usepackage{balance}
\usepackage[bookmarks=false]{hyperref}                     %
\usepackage{courier}                                       % courier font, used in \texttt 
\usepackage{listings}                                      % typeset source code listings
\usepackage[]{algpseudocode}                               % algorithm package
\algrenewcommand\textproc{\texttt}
\makeatletter\let\c@float@type\relax\makeatother
\makeatletter\let\float@addtolists\relax\makeatother
\usepackage{algorithm}
\hypersetup{
    colorlinks = true,
    citecolor  = blue,
    linkcolor  = blue,
    urlcolor   = blue,
}

\makeatletter
\newcommand{\thickhline}{%
	\noalign {\ifnum 0=`}\fi \hrule height 1pt
	\futurelet \reserved@a \@xhline
}

\graphicspath{{./figs/}}

\begin{document}

\title{DeDA: Deep Directed Accumulator}
% \orcidID{1111-2222-3333-4444}

\author{Hang Zhang \inst{1} \Letter \and
Rongguang Wang \inst{2} \and
Renjiu Hu \inst{1} \and
Jinwei Zhang \inst{1} \and
Jiahao Li \inst{1}
}

\authorrunning{H. Zhang et al.}

\institute{Cornell University \and 
University of Pennsylvania \\
\email{hz459@cornell.edu} \\
\url{https://www.tinymilky.com}
}

\maketitle

\begin{abstract}

Chronic active multiple sclerosis lesions, also termed as rim+ lesions, can be characterized by a hyperintense rim at the edge of the lesion on quantitative susceptibility maps.
These rim+ lesions exhibit a geometrically simple structure, where gradients at the lesion edge are radially oriented and a greater magnitude of gradients is observed in contrast to rim- (non rim+) lesions. 
However, recent studies have shown that the identification performance of such lesions remains unsatisfied due to the limited amount of data and high class imbalance.
In this paper, we propose a simple yet effective image processing operation, deep directed accumulator (DeDA), that provides a new perspective for injecting domain-specific inductive biases (priors) into neural networks for rim+ lesion identification.
Given a feature map and a set of sampling grids, DeDA creates and quantizes an accumulator space into finite intervals, and accumulates feature values accordingly.
This DeDA operation is a generalized discrete Radon transform and can also be regarded as a symmetric operation to the grid sampling within the forward-backward neural network framework, the process of which is order-agnostic, and can be efficiently implemented with the native CUDA programming.
Experimental results on a dataset with 177 rim+ and 3986 rim- lesions show that $10.1\%$ of improvement in a partial (false positive rate $<0.1$) area under the receiver operating characteristic curve (pROC AUC) and $10.2\%$ of improvement in an area under the precision recall curve (PR AUC) can be achieved respectively comparing to other state-of-the-art methods.
The source code is available online at \url{https://github.com/tinymilky/DeDA}


\end{abstract}

\section{Introduction}

In the past decade, we have witnessed the significant success of convolutional neural networks (CNNs) being applied to various grid-based \cite{simoncelli2001natural} medical imaging applications such as magnetic resonance imaging (MRI) reconstruction \cite{zhang2021efficient,muckley2021results} and lesion segmentation \cite{zhang2021all,kamnitsas2017efficient}.
% The architecture design of CNNs naturally imposes several desirable inductive biases for grid-based image processing \cite{simoncelli2001natural}, e.g. the translation equivariance through spatially invariant convolution filters \cite{ruderman1993statistics,olshausen1996natural}, the translation invariance \cite{kayhan2020translation} with pooling layers and the locality \cite{lenc2015understanding,lecun1989handwritten}.
% These are most general inductive biases in CNNs and have been proved to be useful for many computer vision applications.
Despite the success of general inductive biases such as translation equivariance \cite{kayhan2020translation} and locality \cite{lenc2015understanding}, medical images represent different diseases and require highly domain-specific knowledge. 
Thus, the question of how to inject domain-specific inductive biases (priors) beyond the general ones into neural networks for medical image processing remains to be answered.

\begin{figure}[!t]
	\centering
	\vspace{-1ex}
        \includegraphics[width=0.98\columnwidth]{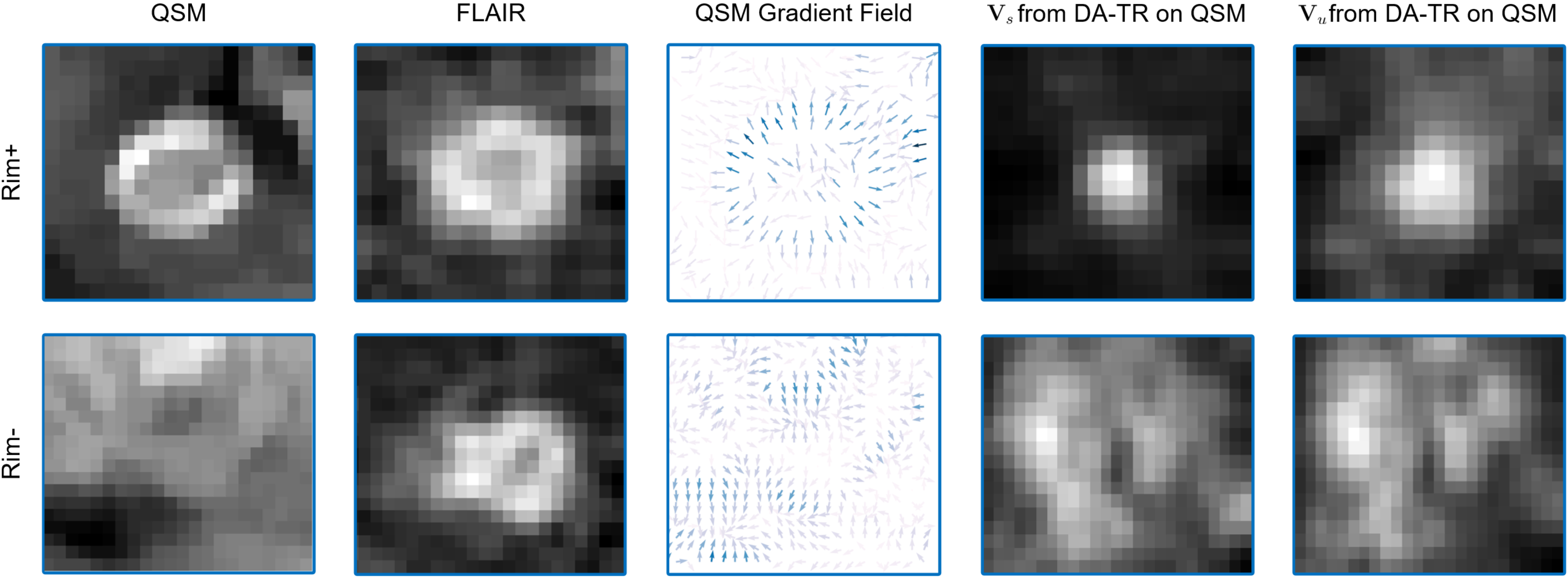}

        \caption{
            A visual example of the difference of a rim+ and a rim- lesion. 
            QSM image patches show the the magnetic susceptibility distribution for the lesions.
            Fluid attenuated inversion recovery (FLAIR) image patches show the exact location of the lesions.
            The gradient filed map of QSM images show gradient vectors normalized to unit vectors (the darker of the blue, the larger of the magnitude of a gradient vector).
            The right two columns are gradient magnitude maps $\mathbf{V}_s$ and QSM value maps $\mathbf{V}_u$ processed by DA-TR (see Section \ref{sec:rim}).
            The rim+ lesion shows structured patterns in the accumulator space by aggregating feature values along gradients; however, the rim- lesion possess no such structures. 
        }
	\label{fig:lesion-grads}
\end{figure}

In this paper, we attempt to answer though tackling the identification problem of a particular type of multiple sclerosis (MS) lesion, called a chronic active lesion (termed as a rim+ lesion).
A rim+ lesion is characterized by an iron-enriched rim of activated macrophages and microglia in histopathology studies \cite{absinta2016persistent,gillen2021qsm,dal2017slow,kaunzner2019quantitative} and are visible with in-vivo quantitative susceptibility mapping (QSM) \cite{wang2015quantitative,wang2017clinical,de2010quantitative} and phase imaging \cite{absinta2016persistent,absinta2013seven} techniques, where these lesions show a paramagnetic hyperintense rim at the edge (see Fig. \ref{fig:lesion-grads}).
Several attempts \cite{barquero2020rimnet,lou2021fully,zhang2022qsmrim} have been made to address the problem, but a clinically reliable one is not yet available.

Considering the limited amount of data and the high class imbalance, it is more desirable to encode priors with domain knowledge into the network explicitly.
It can be seen from the Fig. \ref{fig:lesion-grads} that rim+ lesions differ from rim- (non rim+) lesions in three ways. 
First, rim+ lesions have a hyperintense ring-like structure at the edge of the lesion on QSM.
Second, in rim+ lesions, a greater magnitude of gradients is observed near the lesion edge unlike rim- lesions.
Third, rim+ lesions can be characterized by radially oriented gradients on the edge; however, rim- lesions do not possess such structured orientations.

In this paper, we propose \textbf{De}ep \textbf{D}irected \textbf{A}ccumulator (DeDA), an image processing operation symmetric to the grid sampling within the forward-backward neural network framework to explicitly encode the above prior information into networks.
Given a feature map and a set of sampling grids, DeDA creates and quantizes an accumulator space into finite intervals, and accumulates feature values accordingly.
This DeDA operation can also be regarded as a generalized discrete Radon transform, as it maps values between two discrete functional space through accumulation.
The main contribution of this paper are two folds. 
First, we introduce a simple yet effective method DeDA, a generalized image processing operation, for increasing the representation capacity of neural networks by integrating domain-specific prior information explicitly. 
Second, experimental results on rim+ lesion identification show that $10.1\%$ of improvement in a partial (false positive rate less than $10\%$) area under the receiver operating characteristic curve (pROC AUC) and $10.2\%$ of improvement in an area under the precision recall curve (PR AUC) can be achieved respectively comparing to other state-of-the-art methods.  

\section{Methodology}

A number of signal processing methods such as Fourier transform, Radon transform and Hough transform involve a process of mapping discrete signals from image space to another functional space.
We call this new space accumulator space, because the value of each cell in the new space is a weighted sum of values from the complete set of cells in the original image space. 
One desirable property of the accumulator space for our application is that local convolutions in the accumulator space such as Hough and sinogram space leads to global aggregation of structural features such as lines \cite{lin2020deep,zhao2021deep} in the feature map space, which is beneficial for integrating geometric priors into neural networks.
Unlike attention based methods \cite{wang2018non,zhang2021efficient}, this accumulator space convolution captures long-range information explicitly by direct geometric prior parameterization.

\subsection{Differentiable Directed Accumulation}
\label{sec:deda}

\begin{figure}[!t]
	\centering
        \vspace{-1ex}
	\includegraphics[width=1.0\columnwidth]{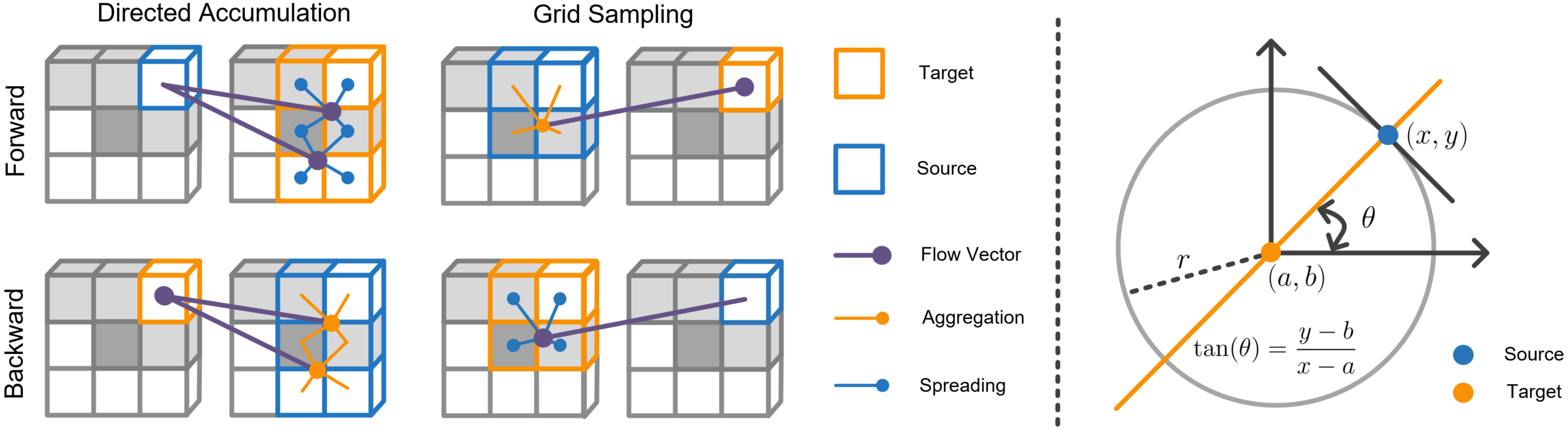}
	\vspace{-1ex}
        \caption{
    Visual illustration of the proposed method. 
    The left panel shows differences between the grid sampling and the proposed DeDA using bilinear sampling kernel.
    The right panel shows the schematic for rim parameterization, where the knowledge of a triple $(x,y,\theta)$ is mapped to a straight line (marked in orange) in the accumulator space.
    % In the forward pass of DeDA, the feature value in the source feature map flows along the given direction into the target cell in the accumulator space, and the final value in the target cell is the sum of all values that flow into it. 
    % In the backward pass, the gradient flows back from the accumulator cell along the opposite direction into the previous value cell.
    % From the figure we can see that the backward pass of DeDA is similar to the grid sampling except the allowance of multiple flow-field grids. 
    % Basically, the proposed DeDA is a symmetric operation to the grid sampling operation, and is further generalized to allow accumulation of values from multiple flow-field grids.
    } 
    \label{fig:deda-gs}
\end{figure}

The process of transforming an image to an accumulator space involves a critical step, directed accumulation (DA), in which a cell from the accumulator space is pointed by multiple cells from the image space.
Fig.~\ref{fig:deda-gs}, Eq.~\eqref{eq:gs} and Eq.~\eqref{eq:deda-back} have shown that this DA operation is a symmetric operation to the grid sampling \cite{jaderberg2015spatial} within the forward-backward learning framework, where the backward pass of DA possesses the same structure as the forward pass of grid sampling if only one sampling grid is given, and vice versa for the forward pass.
In addition, DA is further generalized to allow multiple sampling grids to accumulate values from the source feature map.
Here we briefly review the grid sampling method and then derive the proposed DeDA.
% The most related operation to the directed accumulation is the grid sampling. 
% method that is first introduced in spatial transformer networks \cite{jaderberg2015spatial} which gives neural networks the ability to spatially transform feature maps.
% A successful application of the spatial transformer in the medical field is the deformable image registration \cite{balakrishnan2019voxelmorph}.
% As can seen from Fig. \ref{fig:deda-gs}, the proposed DeDA can be regarded as a generalized symmetric operation of the grid sampling with value accumulation from multiple flow-field grids.
% The proposed DeDA differs from the grid sampling in two folds: 
% 1) The backward pass of DeDA is the same as the forward pass of the grid sampling if only one flow-field is given, and vice versa for the forward pass.
% 2) DeDA is further generalized to allow multiple flow-field grids to accumulate values from the source feature map. 
% Here we briefly review the grid sampling method and then derive the proposed DeDA.

\textbf{Grid Sampling}:
Given a source feature map $\mathbf{U} \in \mathbb{R}^{C\times H \times W}$, a sampling grid $\mathbf{G} \in \mathbb{R}^{2\times H' \times W'}=(\mathbf{G}^x, \mathbf{G}^y)$ specifying pixel locations to read from $\mathbf{U}$, and a kernel function $\mathcal{K}()$ defining the image interpolation, then the output value of a particular position $(i,j)$ at the target feature map $\mathbf{V} \in \mathbb{R}^{C\times H' \times W'}$ can be written as follows:

\begin{equation}
    \mathbf{V}_{ij}^{c} = \sum_n^{H}\sum_m^{W} \mathbf{U}_{nm}^{c}\mathcal{K}(\mathbf{G}_{ij}^x,n)\mathcal{K}(\mathbf{G}_{ij}^y,m),
    \label{eq:gs}
\end{equation}
where the kernel function $\mathcal{K}()$ can be replaced with any other specified kernels, e.g. integer sampling kernel $\delta(\lfloor\mathbf{G}_{ij}^x+0.5\rfloor-n)\cdot \delta(\lfloor\mathbf{G}_{ij}^y+0.5\rfloor-m)$ and bilinear sampling kernel $\text{max}(0,1-|\mathbf{G}_{ij}^x-n|) \cdot \text{max}(0,1-|\mathbf{G}_{ij}^y-m|)$. 
Here $\lfloor x+0.5\rfloor$ rounds $x$ to the nearest integer and $\delta()$ is the Kronecker delta function.
The gradients with respect to $\mathbf{U}$ and $\mathbf{G}$ for back propagation can be defined accordingly \cite{jaderberg2015spatial}.

\textbf{DeDA}:
% 2) Reading values from the source feature map can only be done using the whole cell in the DeDA, while the grid sampling can read weighted values based on the sampling kernel. 
% 3) The DeDA can write weighted values to the target feature map, while writing values to the target feature map can only done using the whole cell in the grid sampling.
Given a source feature map $\mathbf{U} \in \mathbb{R}^{C\times H \times W}$, a target feature map $\mathbf{V} \in \mathbb{R}^{C\times H' \times W'}$, a set of sampling grids $\mathcal{G} = \{\mathbf{G}[k] \in \mathbb{R}^{2\times H \times W}=(\mathbf{G}^x[k], \mathbf{G}^y[k])~|~k \in \mathbb{Z}^+, 1 \leq k \leq N \}$ ($N\geq 1$ is the number of grids), and a kernel function $\mathcal{K}()$, the output value of a particular position $(i,j)$ at the target feature map $\mathbf{V}$ can be written as follows:
\vspace{-1ex}
\begin{equation}
    \mathbf{V}_{ij}^{c} = \sum_k^{N}\sum_n^{H}\sum_m^{W} \mathbf{U}_{nm}^{c}\mathcal{K}(\mathbf{G}_{nm}^x[k],i)\mathcal{K}(\mathbf{G}_{nm}^y[k],j).
    \label{eq:deda}
    \vspace{-1ex}
\end{equation}
It is worth noting that the spatial dimension of the grid $\mathbf{G}[k]$ should be the same as that of $\mathbf{U}$, but the first dimension of $\mathbf{G}[k]$ can be an arbitrary number as long as it aligns with the number of spatial dimensions of $\mathbf{V}$, e.g. if given $\mathbf{U} \in \mathbb{R}^{H \times W}$ and $\mathbf{G}[k] \in \mathbb{R}^{3\times H \times W}$, it is expected that $\mathbf{V} \in \mathbb{R}^{H' \times W' \times D'}$.
Basically, the DeDA operation in Eq.~\eqref{eq:deda} performs a function mapping by $\mathcal{D}:(\mathbf{U},\mathcal{G};\mathcal{K})\rightarrow \mathbf{V}$, where $\mathcal{K}$ is the sampling kernel.
For simplicity, function $\mathcal{D}()$ will be used to denote the DeDA forward for the rest of the paper.

To allow back propagation for training networks with DeDA, the gradients with respect to $\mathbf{U}$ are derived using the chain rule as follows:
\vspace{-1ex}
\begin{equation}
\label{eq:deda-back}
    \dfrac{\partial\mathcal{L}}{\partial \mathbf{V}_{nm}^{c}}\dfrac{\partial\mathbf{V}_{nm}^{c}}{\partial\mathbf{U}_{ij}^{c}} = \sum_k^{N}\sum_n^{H'}\sum_m^{W'} \mathbf{A}_{nm}^{c}\mathcal{K}(\mathbf{G}_{ij}^x[k],n)\mathcal{K}(\mathbf{G}_{ij}^y[k],m),
    \vspace{-1ex}
\end{equation}
where $\mathbf{A}$ is the gradient tensor with respect to $\mathbf{V}$. 
We can see that the structure of Eq. \eqref{eq:deda-back} reduces to Eq. \eqref{eq:gs} 
by setting $N=1$, meaning that DeDA is a symmetric operation to grid sampling.
Note that both the forward and backward pass of DeDA transforms each channel $c$ in an identical way and thus without loss of generality, the feature map is denoted with spatial dimensions only for the rest of the paper.

% To guide the generation of flow-field grid coordinates and allow loss gradients flow back to corresponding network weights, we further derive the the gradients with respect to $\mathbf{G}^x$ as follows:
% \begin{equation}
% % \resizebox{1\columnwidth}{!}{
%      \dfrac{\partial\mathcal{L}}{\partial \mathbf{V}_{nm}^{c}}\dfrac{\partial\mathbf{V}_{nm}^{c}}{\mathbf{G}_{ij}^x} = \sum_k^{N}\sum_n^{H'}\sum_m^{W'} \mathbf{A}_{nm}^{c}\mathbf{U}_{ij}^{c}\mathcal{K}(\mathbf{G}_{ij}^y[k],m) t,
%      \label{eq:deda-back-coord}
% % }
% \end{equation}
% where the value of $t$ depends on the sampling kernel. 
% Taking bilinear sampling kernel as an example, where $\mathcal{K}(\mathbf{G}_{ij}^x[k],n)=\text{max}(0,1-|\mathbf{G}_{ij}^x[k]-n|)$, then we can describe $t$ in Eq.~\eqref{eq:deda-back-coord} as follows:
% \begin{equation}
%     t = \begin{cases}
%       0 & \text{if} ~~ |\mathbf{G}_{ij}^x[k]-n|\geqslant 1\\
%       1 & \text{if} ~~ \mathbf{G}_{ij}^x[k] \leqslant n \\
%      -1 & \text{if} ~~ \mathbf{G}_{ij}^x[k] > n,
%     \end{cases}   
% \end{equation}
% where sub-gradients are used to address the discontinuity in the sampling kernel function. 
% Similar derivation for $ \dfrac{\partial\mathcal{L}}{\partial \mathbf{V}_{nm}^{c}}\dfrac{\partial\mathbf{V}_{nm}^{c}}{\mathbf{G}_{ij}^y}$ can be obtained as well.

\subsection{DeDA-based Transformation Layer for Rim Parameterization}
\label{sec:rim}

In this section, we derive DeDA-based transformation and its convolution layers for rim parameterization.
As shown in Fig.~\ref{fig:lesion-grads}, a rim+ lesion can be characterized by a hyperintense rim at the lesion edge on QSM and differs from a rim- lesion in both image intensities and gradients at the edge. 
% be parameterized as follows:
% \begin{equation}
%     (x-a)^2 + (y-b)^2 = r^2,
%     \label{eq:hct}
% \end{equation}
% where $(a,b)$ and $r$ are parameters describing the center coordinates and the radius of the circle.
% The HCT uses Eq.~\eqref{eq:hct} to map each image pixel $(x,y)$ into a inverted right angled cone with its apex at $(x,y,0)$ embedded in a 3D accumulator space.
% To simplify the accumulator space, we follow the adaptive HCT \cite{illingworth1987adaptive} and parameterize a 2D accumulator space with $(a,b)$ only, in which we impose a mild constraint that all vectors normal to the circle edge should intersect at the circle center.
To account for both image intensities and gradients, the rim is parameterized as $\tan(\theta) = \dfrac{y-b}{x-a}$,
where $(a, b)$ are parameters of coordinates for the rim center in the accumulator space and $\theta$ represents the gradient direction at $(x,y)$ in the image space.
% In our application, the normal vector of each pixel $(x,y)$ can be obtained by local image gradients, leading to a $\theta$ representing the gradient direction.
As can be seen from the right panel of Fig. \ref{fig:lesion-grads}, mapping a single $(x,y,\theta)$ to the accumulator space produces a straight line, and thus coordinates of the rim center can be identified by the intersection of many of these lines.
% \begin{equation}
%     \tan(\theta) = \dfrac{y-b}{x-a},
%     % b = \tan(\theta)a+ (y-x\tan(\theta)),
%     \label{eq:s-hct}
% \end{equation}

\textbf{DeDA Transformation of the Rim}:
Given a source feature map $\mathbf{U} \in \mathbb{R}^{H \times W}$, the magnitude of image gradients can be obtained as follows $\mathbf{S} = \sqrt{\mathbf{U}_x\odot\mathbf{U}_x + \mathbf{U}_y\odot\mathbf{U}_y}$, where $\odot$ denotes the Hadamard product, $\mathbf{U}_x = \dfrac{\partial \mathbf{U}}{\partial x}$, and $\mathbf{U}_y = \dfrac{\partial \mathbf{U}}{\partial y}$.
The image gradient tensor $\mathbf{U}_x$ and $\mathbf{U}_y$ can be efficiently computed using convolution kernels such as the Sobel operator.
Normalized gradients can be obtained by $\hat{\mathbf{U}}_x = \dfrac{\mathbf{U}_x}{\mathbf{S}+\epsilon}$ and $\hat{\mathbf{U}}_y = \dfrac{\mathbf{U}_y}{\mathbf{S}+\epsilon}$, where $\epsilon$ is a small real value to avoid zero denominator.
The mesh grids of $\mathbf{U}$ are denoted as $\mathbf{M}_x$ (value range: $(0,H-1)$) and $\mathbf{M}_y$ (value range: $(0,W-1)$).
We can then generate a set of sampling grids as follows:
\begin{equation}
    \mathcal{G} = \{\mathbf{G}[k]=(\mathbf{G}^x[k], \mathbf{G}^y[k])~|~k \in \mathbb{Z}^+, 1 \leq k \leq N \},
\end{equation}
where $\mathbf{G}[k] \in \mathbb{R}^{2\times H \times W}$, $\mathbf{G}^x[k]=k\hat{\mathbf{U}}_x+\mathbf{M}_x$, $\mathbf{G}^y[k]=k\hat{\mathbf{U}}_y+\mathbf{M}_y$, and $N=\text{max}(H,W)$.
% Let $\mathcal{G}^{-}$ be the sampling grid set having gradients in the opposite direction, where  $\mathbf{G}^x[k]^{-}=-k\hat{\mathbf{U}}_x+\mathbf{M}_x$, $\mathbf{G}^y[k]^{-}=-k\hat{\mathbf{U}}_y+\mathbf{M}_y$.
Now the DeDA-based transformation of Rim (DA-TR) can be formulated as $\mathbf{V}_s = \mathcal{D}(\mathbf{S},\mathcal{G};\mathcal{K})$  and $\mathbf{V}_u = \mathcal{D}(\mathbf{U},\mathcal{G};\mathcal{K})$, where the integer sampling kernel is used.
It is worth noting that feature and gradient magnitude values are accumulated separately due to differences of image intensity and gradients between rim+ and rim- lesions (see Fig.~\ref{fig:lesion-grads}).
% To add symmetric information of rim to the accumulator space, an additional sampling grid set $\mathcal{G}^{-}$ is applied.

\begin{figure}[!t]
	\centering
	\vspace{-1ex}
        \includegraphics[width=1.0\columnwidth]{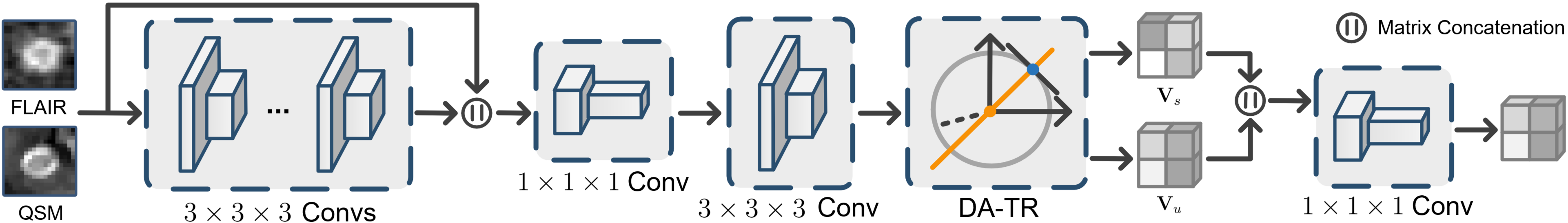}
	\vspace{-1ex}
        \caption{
            Schematic of the network layer for DA-TR.
            Conv denotes a convolutional layer, and each of these layers consists a $3\times3\times3$ or $1\times1\times1$ convolution, a batch normalization, and a ReLU activation.
        }
	\label{fig:da-rim}
\end{figure}
\textbf{Network Layer for DA-TR}:
To gain more representation ability and capture long-range contextual information, DA-TR is applied to both intermediate feature maps and original images. 
As can be seen from Fig.~\ref{fig:da-rim}, image patches of lesions are processed through a set of convolutional layers with each consisting of a $3\times3\times3$ or $1\times1\times1$ convolution, a batch normalization \cite{ioffe2015batch} and a ReLU activation function, followed by a DA-TR layer and a $1\times1\times1$ convolutional layer.
The first $1\times1\times1$ conv layer is used to fuse feature maps and original image patches for better feature embedding, and the second one is used to fuse DeDA transformed gradient magnitude maps $\mathbf{V}_s$ and feature maps $\mathbf{V}_u$. 
It is worth noting that only in-plane rims are observed, and thus the DA-TR is performed on the 2D feature map slices along the axial direction.

\section{Experiments and Results}

For fair and consistent comparison, the dataset applied in the previous work \cite{zhang2022qsmrim} was asked for and used to demonstrate the performance of the proposed DeDA-based rim parameterization DA-TR.
A total of 172 subjects were included in the dataset, and 177 lesions were identified as rim+ lesions and 3986 lesions were identified as rim- lesions, please refer to \cite{zhang2022qsmrim} for more details about the image acquisition and pre-processing.

\subsection{Comparator Methods and Implementation Details}

\textbf{Comparator Methods}:
Three methods have been developed so far for rim+ lesion identification, of which APRL \cite{lou2021fully} and RimNet \cite{barquero2020rimnet} are on phase imaging and QSMRim-Net \cite{zhang2022qsmrim} is on QSM.
In comparison with these methods, we use QSM along with T2-FLAIR images as the network inputs for RimNet and QSMRimNet, and use the QSM image to extract first-order radiomic features for APRL.
Furthermore, we also applied residual networks (ResNet) \cite{he2016deep}, vision transformer (ViT)\cite{dosovitskiy2020image}, Swin transformer \cite{liu2021swin}, and Nested transformer \cite{zhang2022nested} as backbone architecture for our application, and determined that ResNet with 18 convolution layers works the best.
Transformer-based networks with fewer inductive biases rely heavily on the use of a large training dataset or depends strongly on the feature reuse \cite{matsoukas2022makes}, as a result, these networks as well as CNNs with deeper structures are prone to overfit small datasets.
Therefore, integrating proper priors into networks is crucial for rim+ lesion identification.

\textbf{Implementation Details}:
A stratified five-fold cross-validation procedure was applied to train and validate the performance, and all experiments including ablation study were carried out within this setting.
Each lesion was cropped into patches with a fixed size of $32\times32\times8$ voxels.
Random flipping, random affine transformation and random Gaussian blurring were used to augment our data.
% Our network was implemented in version 3.7 of Python with version 1.9.0 of PyTorch library \cite{paszke2019pytorch} on a computer equipped with two Nvidia Titan XP GPUs. 
% Particularly, the DeDA operation in Eq.~\ref{eq:deda} was implemented in C++ with CUDA 11.1.
More details of the training procedure can be found out in the appendix.

\begin{figure}[!t]
	\centering
        \vspace{-1ex}
	\subfloat[Pearson's Corr]{\includegraphics[width=.26\columnwidth]{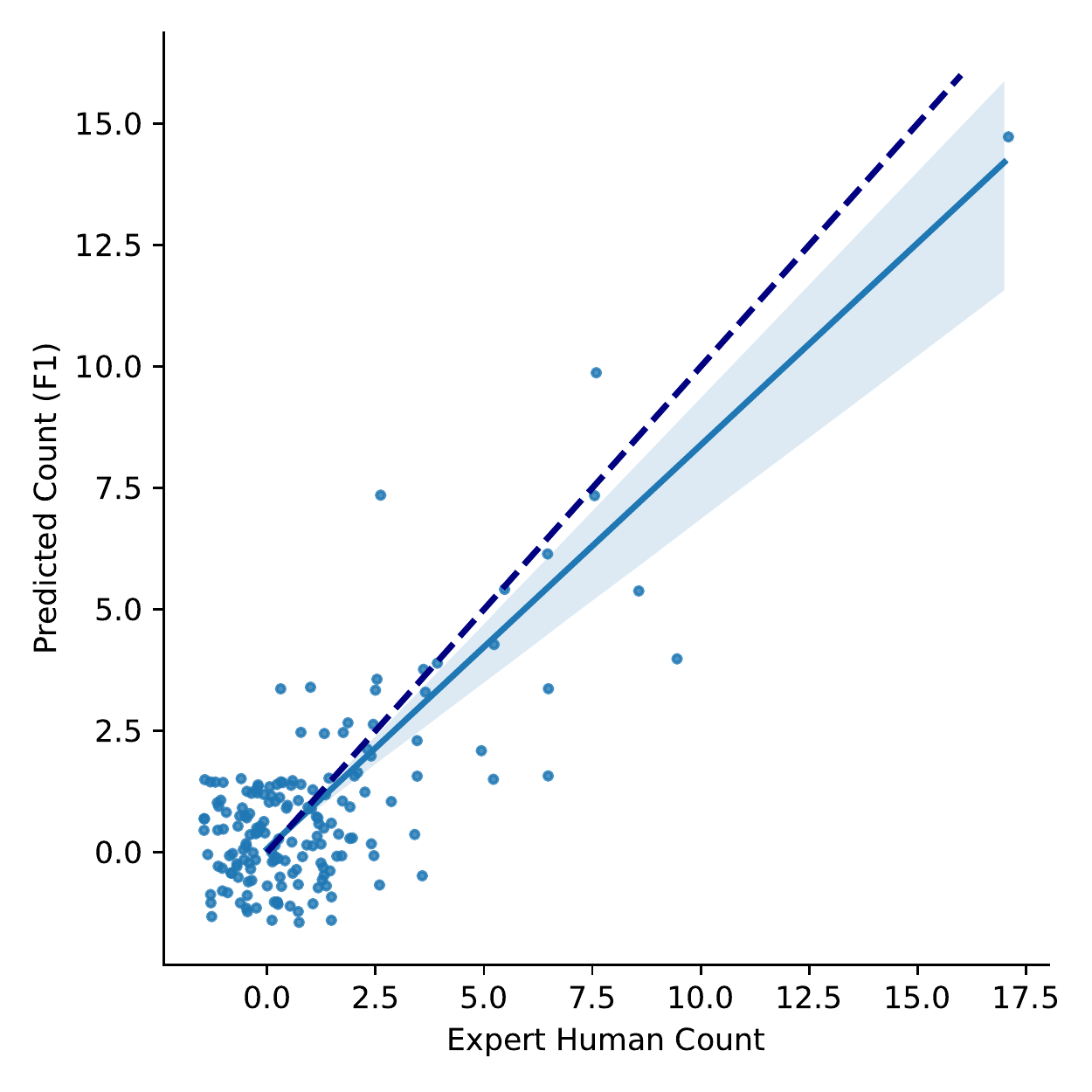}
    \label{fig:subject-wise}}
	\subfloat[pROC Curves]{\includegraphics[width=.37\columnwidth]{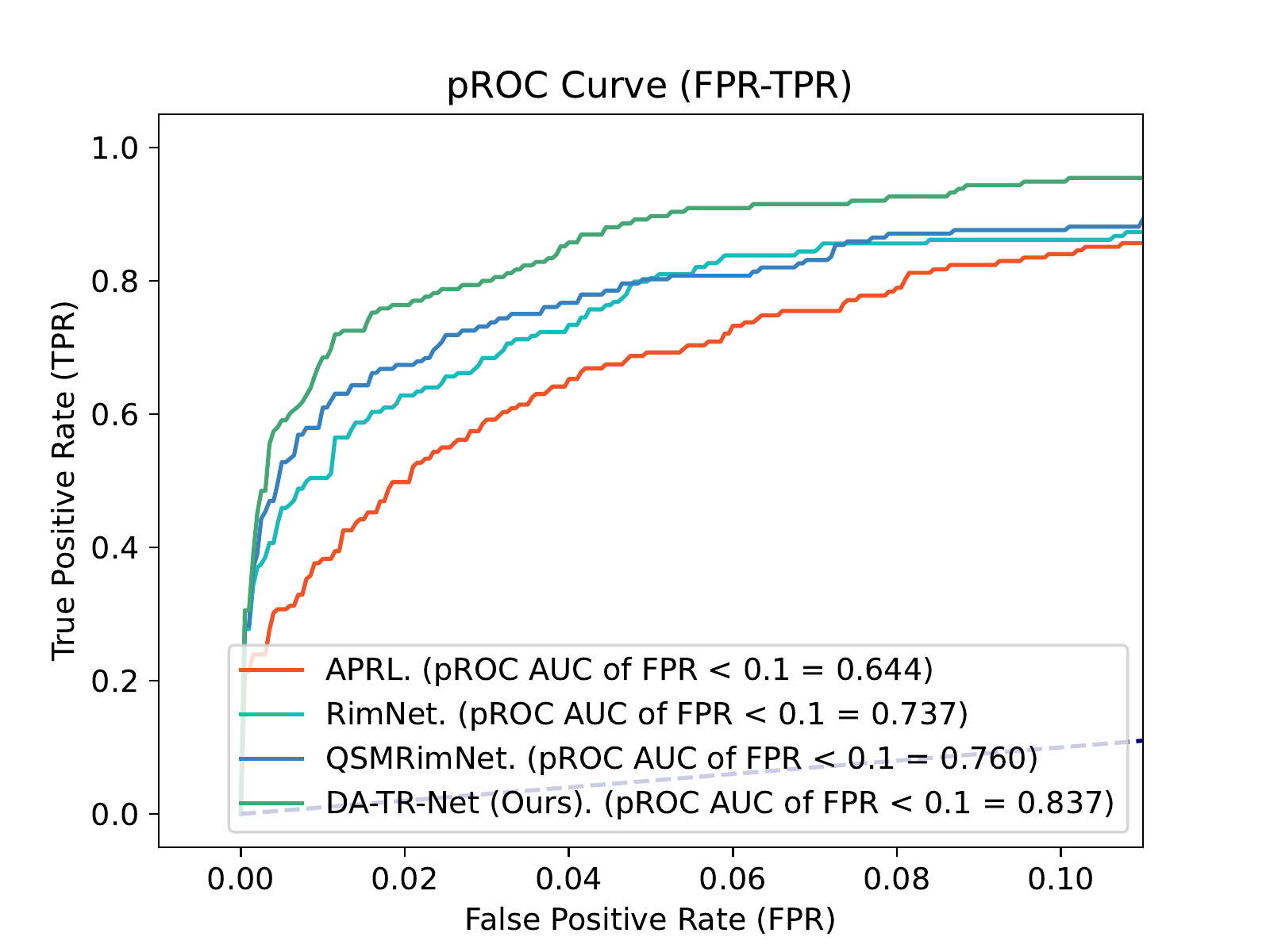}
    \label{fig:pROC}} 
	\subfloat[PR Curves]{\includegraphics[width=.37\columnwidth]{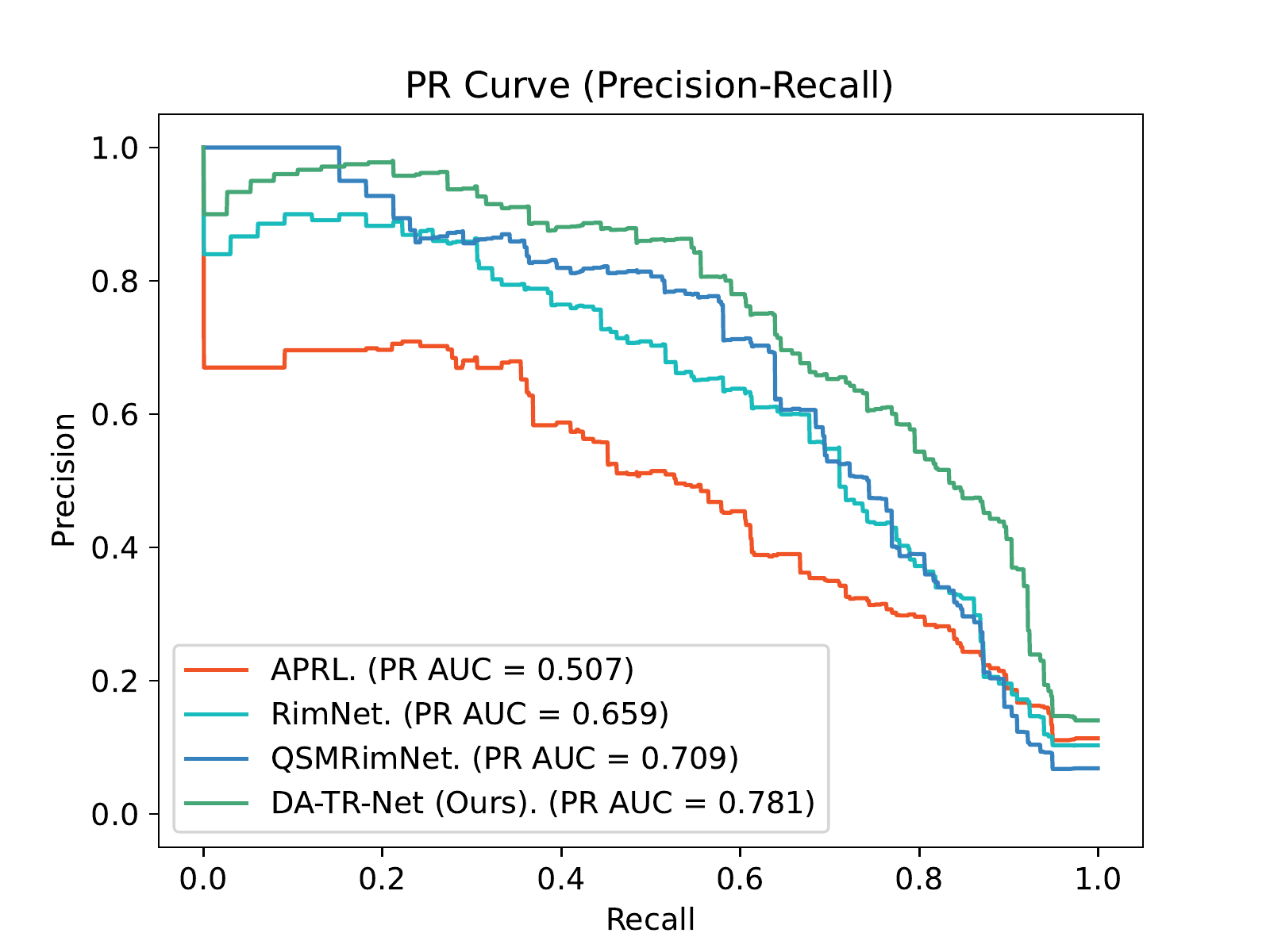}
    \label{fig:PR}}
    \vspace{-1ex}
	\caption{
    	The predicted count of rim + lesions from DA-TR-Net versus the expert human count is shown in (a), where points in the plot have been jittered for better visualization. 
            The pROC and PR curves for the proposed and other comparator methods are shown in (b) and (c), where AUC denotes the area under the curve.
	\label{fig:curves}
        }
        \vspace{-2ex}
\end{figure}

\begin{table}[!b]

\caption{ Results of the proposed and other methods using a stratified five-fold cross-validation scheme.
The best performing metric is bolded.
}
\vspace{-1ex}
\label{tab:overall}
%\vskip 0.15in
\begin{center}
\resizebox{1.\columnwidth}{!}{
\begin{tabular}{ lcccccccccc}
\hline
\hline
Method	&Accuracy	&$F_1$ 	&Sensitivity	&Specificity	&Precision	&ROC AUC	&pROC AUC &PR AUC & $\rho$ (95\%CI) & MSE \\
\hline
APRL \cite{lou2021fully}	&0.954	&0.538	&0.627	&0.969	&0.470	&0.940	&0.644	&0.507	&0.68 (0.59,0.75)	&3.16	\\
RimNet \cite{barquero2020rimnet}	&0.970	&0.650	&0.655	&0.984	&0.644	&0.950	&0.737	&0.659	&0.75 (0.67,0.81)	&2.41	\\
QSMRimNet \cite{zhang2022qsmrim}	&0.977	&0.711	&0.667	&0.991	&0.761	&0.939	&0.760	&0.709	&0.89 (0.86,0.92)	&1.00	\\
DA-TR-Net (Ours)	&\bf{0.980}	&\bf{0.750}	&\bf{0.712}	&\bf{0.992}	&\bf{0.792}	&\bf{0.975}	&\bf{0.837}	&\bf{0.781}	&\bf{0.93(0.90,0.95)}	&\bf{0.69}	\\

\hline
\hline
\end{tabular}
}
\end{center}
\vspace{-1ex}
\end{table}

\subsection{Results and Ablation Study}

% The Adam algorithm \cite{kingma2014adam} with an initial learning rate of $1e-4$ and a multi-step learning rate scheduler with rate halving at 50\%, 70\%, and 90\% of the total epochs, were used for training. 
% A mini-batch size of 32 was used for training, and training was stopped after 50 epochs. 
% We used three random seeds to train three models for each fold and the final prediction result was determined by ensembing of logits from three models, followed by a Sigmoid function to produce probability. 
% A sensitivity analysis was performed and found out that three random seeds were well in terms of the computational cost versus the performance gain as a balance of all performance metrics.

\textbf{Lesion-wise Results}: 
To evaluate the performance of each method and produce clinically relevant results, pROC curves with false positive rates (FPRs) in the range of $(0,0.1)$ and PR curves of the different validation folds were interpolated using piece-wise constant interpolation and averaged to show the overall performance at the lesion level. 
For each curve, AUC was computed directly from the interpolated and averaged curves. 
The binary indicators of rim+/rim- lesions were generated by thresholding the model probabilities to maximize the $F_1$ score, where $F_1=2\cdot\dfrac{precision\cdot sensitivity}{precision+sensitivity}$. 
In addition, accuracy, F1 score, sensitivity, specificity, and precision were used to characterize the performance of each method. 
Table \ref{tab:overall} and Fig. \ref{fig:curves} show the lesion-wise performance metrics of the proposed methods in comparison with the other methods. 
DA-TR-Net outperformed the other competitors in all evaluation metrics. 
With a slightly higher overall accuracy and specificity with other methods, DA-TR-Net resulted in a 5.5\%, 15.4\% and 39.4\% improvement in $F_1$ score, 10.1\%, 13.6\% and 30.0\% improvement in pROC (FPR$<0.1$) AUC, and 10.2\%, 18.5\% and 54.0\% improvement in PR AUC compared to QSMRimNet, RimNet and APRL, respectively. 

\textbf{Subject-wise Results}: We also evaluated the performance at the subject-level. 
Pearson’s correlation coefficient was used to measure the correlation model predicted count and human expert count.
Mean Squared Error (MSE) was also used to measure the averaged accuracy for the model predicted count.
Fig. \ref{fig:subject-wise} shows the scatter-plot for the predicted count v.s. the human expert count, along with the identity line, and the Pearson’s correlation coefficient ($\rho$) for DA-TR-Net was $\rho=0.93 (95\%CI:0.90,0.95)$
As can be seen from Table \ref{tab:overall}, the Pearson’s correlations and MSE for the proposed DA-TR-Net was found higher than other competitors.
This demonstrates that the performance of DA-TR-Net at the subject-level is statistically significantly higher than that of APRL, Rim-Net, and QSMRim-Net.

\textbf{Ablation Study}:
We conducted an ablation study to investigate the effects of each component accompanied with DA-TR.
First, we examined the effects of applying the proposed DA-TR to the latent feature maps and raw images.
Second, we examined the effects of using $\mathbf{V}_u$ and $\mathbf{V}_s$, because rim+ lesions differ from rim- lesions in both gradient magnitudes and values at the edge of the lesion.
We then investigated how multi-radius rim parameterization can affect the results, as the size of rim+ lesions vary greatly with a radius from 5 to 15 among different subjects. 
Results from models $\#1$, $\#2$ and $\#4$ show that the rim parametrization DA-TR is useful for rim+ identification, and DA-TR used in the latent feature map space performs even better.
Comparing model $\#3$ and $\#4$, one can see that accumulating both gradient magnitudes and feature values is beneficial.
The consistent performance improvement from model $\#4$ to $\#5$ and from model $\#5$ to $\#6$ has demonstrated the effectiveness of applying multi-radius rim parameterization.
More results on backbone networks can be found in the appendix.

\begin{table}[!t]

\caption{ Ablation study on the effects for each component in DA-TR. 
Multiple check marks for sets of $N$ denote the union of the checked sets.
Pre-Convs denotes a convolution block with six $3\times 3 \times 3$ convolution layers.  
}

\label{tab:ablation}
%\vskip 0.15in
\begin{center}
\resizebox{1\columnwidth}{!}{
\begin{tabular}{ccccccccccc}
\hline
\hline
\#	&Pre-Convs	&$\mathbf{V}_u$	&$\mathbf{V}_s$	&$N \in \{5,7,9\}$	&$N \in \{11,13\}$	&$N \in \{15\}$	&$F_1$	&ROC AUC	&pROC AUC (FPR$<0.1$)	&PR AUC	\\

\hline
1	&$\times$	&$\times$	&$\times$	&$\times$	&$\times$	&$\times$	&0.685	&0.945	&0.753	&0.689	\\
2	&$\times$	&$\checkmark$	&$\checkmark$	&$\times$	&$\times$	&$\checkmark$	&0.701	&0.971	&0.790	&0.720	\\
3	&$\checkmark$	&$\checkmark$	&$\times$	&$\times$	&$\times$	&$\checkmark$	&0.703	&0.967	&0.795	&0.714	\\
4	&$\checkmark$	&$\checkmark$	&$\checkmark$	&$\times$	&$\times$	&$\checkmark$	&0.702	&0.976	&0.817	&0.736	\\
5	&$\checkmark$	&$\checkmark$	&$\checkmark$	&$\times$	&$\checkmark$	&$\checkmark$	&0.727	&0.975	&0.825	&0.743	\\
6	&$\checkmark$	&$\checkmark$	&$\checkmark$	&$\checkmark$	&$\checkmark$	&$\checkmark$	&0.750	&0.975	&0.837	&0.781	\\

\hline
\hline
\end{tabular}
}
\end{center}

\end{table}

\section{Conclusions}

% In this paper, we proposed DeDA, an image processing operation symmetric to the grid sampling within the forward-backward neural network learning framework to explicitly encode the prior information of rim into networks.
% This DeDA operation can also be regarded as a generalized discrete Radon transform, as it maps values between two discrete functional space through accumulation.
% Experimental results show that our proposed methods outperform other state-of-the-art methods in all evaluation metrics by a significant margin.  
% In addition, DeDA is a general image processing and can be applied in other applications such as Hough transform, bilateral grid and Polar transform.
% We believe DeDA can benefit more medical applications in the future.

We present DeDA, an image processing operation that helps parameterize rim and effectively incorporates prior information into networks through a value accumulation process. 
The experimental results demonstrate that DeDA surpasses existing state-of-the-art methods in all evaluation metrics by a significant margin. 
Furthermore, DeDA's versatility extends beyond lesion identification and can be applied in other image processing applications such as Hough transform, bilateral grid, and Polar transform. 
We are excited about the potential of DeDA to advance numerous medical applications and other image processing tasks.

\textbf{Acknowledgement}: The database was approved by the local Institutional Review Board and written informed consent was obtained from all patients prior to their entry into the database. 
We would like to thank folks from Weill Cornell for sharing the data used in this paper.

\newpage

\section{Appendix}

\subsection{Preliminaries on Quantitative Susceptibility Maps (QSM) and Rim+ Lesions}

QSM is an MRI imaging technique that can measure the underlying tissue apparent magnetic susceptibility \cite{stuber2016iron,de2010quantitative}, quantifying specific biomarkers such as iron that are independent of imaging parameters and field strength \cite{deh2015reproducibility}.
The forward model of generating magnetic field from susceptibility maps with additive noise is a spatial convolutional process and can be described as the following:
\begin{equation}
    b = \chi * d + n,
    \label{eq:qsm}
\end{equation}
where $b$ is the magnetic field, $\chi$ is the tissue susceptibility, $d$ is the dipole convolution kernel, and $n$ is the additive measurement noise.
Given $b$ and $d$, QSM recovers $\chi$ from solving the ill-posed dipole inversion problem \cite{wang2015quantitative} in Eq.~\eqref{eq:qsm}.
Rim + lesions are characterized by a paramagnetic rim with iron deposited at the edge of the lesion.
QSM is sensitive to such magnetic susceptibility changes and provides consistent measurements of the susceptibility value of the rim across patients and scanners, which is beneficial for a machine learning model such as deep neural network to learn patterns of rim+ versus rim- lesions.

Recent studies have shown that patients with the presence of rim+ lesions are associated with a more severe disease course \cite{Marcille2022,absinta2019association} and there are growing interests in using these lesions as an imaging biomarker.
However, recent a few methods \cite{zhang2022qsmrim,barquero2020rimnet,lou2021fully} developed for rim+ lesion identification are not satisfactory, partially because CNNs with deeper structures or transformer based neural networks \cite{dosovitskiy2020image,liu2021swin} injected with fewer inductive biases demand very large datasets or feature reuse \cite{matsoukas2022makes}, and a method with domain-specific priors describing rims is lacking.
% more powerful only when trained through massive data sets.
% In rim+ lesion identification, former methods \cite{zhang2022qsmrim,barquero2020rimnet,lou2021fully} generalize poorly due to the high data imbalance and insufficient amount of data.
We first brief a set of methods that integrate priors with value accumulation, and then describe recent neural networks using other priors.

\subsection{Related Works}

There are quite a few classic methods involving a process of accumulating feature values, and examples include not limited to histogram of gradients \cite{dalal2005histograms}, local binary pattern \cite{guo2010completed} and polar transformation \cite{esteves2018polar}. 
Hough transform (HT) \cite{osti4746348} and its subsequent variants or improvements are most commonly seen methods that take advantages of the value accumulation process.
% HT is originally devised for straight line detection \cite{osti4746348}, and then generalized for arbitrary shape detection \cite{ballard1981generalizing,duda1972use}.
% HT can efficiently accumulate evidence from local image features for all possible target shapes and is sensitive to the presence of only part of a shape.
The key idea is the accumulation process that maps local image features to an application-specific accumulator space such as line parameterization using polar coordinates \cite{lin2020deep,zhao2021deep}.
We will brief methods using accumulator space voting for obtaining peak responses, and then describe methods utilizes accumulator space priors with convolutions, followed by more general networks with prior information.
% and leave the methods with the accumulation by prediction in the supplementary materials.
 
\subsubsection{Accumulation by Prediction}
Leibe \emph{et al.} \cite{leibe2006interleaving} proposes a probabilistic Hough voting model for object detection, and the parameters of which are further optimized by a max-margin constraint \cite{maji2009object} for better performance. 
LV-Metric \cite{codella2008left} uses HT to segment left ventricle in cine MRI images.
Deep Voting \cite{xie2015deep} generates votes via neural networks for nucleus localization in microscopy images.
Hough-CNN \cite{milletari2017hough} utilizes Hough voting to improve the performance of MRI and ultrasound image segmentation.
The state-of-the-art performance for object detection in 3D point clouds has also been achieved by network-predicted votes \cite{qi2019deep,qi2020imvotenet}.
Memory U-Net \cite{zhang2021memory} generates Hough votes using CNNs for lesion instance segmentation.
This line of works mainly uses classical or learning models to generate votes for object detection or segmentation, leaving much room for exploiting the global prior information from the accumulator space.

\subsubsection{Accumulator Space Convolution}

Local convolution filters applied to the accumulator space leads to aggregation of structural features such as lines \cite{lin2020deep,zhao2021deep} and circles in the feature map space, which is beneficial for integrating priors into networks.
Unlike attention based methods \cite{wang2018non,zhang2021efficient}, this accumulator space convolution captures long-range information explicitly by direct prior parameterization.
% the following works further utilizes the global prior information injected by the accumulator space and apply convolution in the new space to learn structured information such as lines \cite{lin2020deep,zhao2021deep}, or circles in our work.
Lin \emph{et al.} \cite{lin2020deep} uses HT as a global prior for line parameterization to segment straight lines, and Zhao \emph{et al.} \cite{zhao2021deep} incorporates accumulator space into the loss function for improved semantic line detection.
Interestingly, semantic correspondence detection for both 3D point clouds \cite{lee2021deep} and 2D images \cite{min2019hyperpixel} have also been improved by performing convolutions in the accumulator space.
Zhao \emph{et al.} \cite{zhao20223d} combines Manhattan world assumption and latent features from CNNs using HT for 3D room layout estimation.
Our work generalizes the key step, i.e. the value accumulation along a given direction, from pioneer works, and makes possible incorporating a wide range of priors from classical image processing methods into neural networks.

\subsubsection{Network with Priors}

Training deep networks usually demands very large datasets \cite{deng2009imagenet,lin2014microsoft}, which is difficult and even impossible for resource-limited clinical applications. 
% The priors encoding domain-specific knowledge can reduce the amount of data needed for network training or improve the task performance, providing flexibility to resource-limited clinical applications.
The priors encoding domain-specific knowledge has shed light and provides flexibility to resource- and data-limited clinical applications.
The distance transformation mapping \cite{ma2020distance} and spatial information encoding \cite{liu2018intriguing} have been proven successful in developing a variety of edge-aware loss functions \cite{kervadec2019boundary,9434085,karimi2019reducing}, network layers with anatomical coordinates \cite{zhang2021all} as the prior information, and spatially covariant network weight generation \cite{zhang2023spatially}, improving the performance of medical image segmentation.
Ill-posed medical image reconstruction problems rely heavily on carefully designed regularization priors and inserting the prior knowledge of a physical equation \cite{o2017introduction} describing the inverse problem has been proven very effective.
Typical works include but not limited to using deep unfolding network \cite{hershey2014deep} to approximate the forward physical model \cite{8434321,hammernik2018learning}, inserting ADMM solver \cite{boyd2011distributed} into the network training phase for compressed sensing MRI \cite{NIPS2016_1679091c}, and tuning network weights with fidelity-imposed loss function \cite{zhang2020fidelity,jiang2020neural}, acceleration of MRI acquisition \cite{zhang2023laro}, reconstruction of brain quantitative susceptibility for oxygen extraction fraction mapping \cite{cho2022qq} and myelin water fraction mapping \cite{kim2022subsecond}.
In addition, there are other works injecting physics knowledge into learning models to improve performance for systems such as lithography \cite{zhang2016enabling,zhang2017bilinear,jiang2019fast}, thermal conductance cooling \cite{hu2022thermal}, and massive machine-type communications \cite{teng2020accumulated}.     
Polar or log polar features have also been widely used in various applications including but not limited to modulation classification \cite{teng2020accumulated}, rotation- and scale-invariant polar transformer network \cite{esteves2018polar}, general object detection \cite{xie2020polarmask,xu2019explicit,park2022eigencontours}, correspondence matching \cite{ebel2019beyond}, and cell detection \cite{schmidt2018cell} and segmentation \cite{stringer2021cellpose}.
Our work DeDA, a simple operation symmetric to the grid sampling, makes it simple and fast to integrate customized priors such as rim parametrization into neural networks.
% Typical works include but not limited to using deep unfolding network \cite{hershey2014deep} to approximate the forward physical model \cite{8434321,hammernik2018learning}, inserting ADMM solver \cite{boyd2011distributed} into the network training phase for compressed sensing MRI \cite{NIPS2016_1679091c}, tuning network weights with fidelity loss \cite{zhang2020bayesian,zhang2020fidelity,zhang2021probabilistic,zhang2021temporal}, accelerating MRI acquisition \cite{zhang2022laro}, and reconstructing brain quantitative susceptibility for oxygen extraction fraction mapping \cite{cho2022qq} and myelin water fraction mapping \cite{kim2022subsecond}. 
% Polar or log polar features have also been widely used in various applications including but not limited modulation classification \cite{teng2020accumulated}, rotation- and scale-invariant polar transformer network \cite{esteves2018polar}, general object detection \cite{xie2020polarmask,xu2019explicit,park2022eigencontours}, correspondence matching \cite{ebel2019beyond}, and cell detection \cite{schmidt2018cell} and segmentation \cite{stringer2021cellpose}.

\subsection{Experiments and Results}

\subsubsection{More Details on the Network Training}

We applied a stratified five-fold cross-validation procedure to train and validate the performance of our methods and the other methods. 
The stratified procedure was performed to balance the number of rim+ lesions in each of the five folds. 
We first grouped subjects into four groups, where the first group contained subjects with no rim + lesion, the second subjects with 1–3 rim + lesions, the third subjects with 4–6 rim + lesions, and the fourth subjects with more than 6 rim + lesions. 
The data was then randomly split into the five folds within each of these groups. 
All experiments were conducted within this stratified five-fold cross validation setting.

Our network was implemented in version 3.7 of Python with version 1.9.0 of PyTorch library \cite{paszke2019pytorch} on a computer equipped with two Nvidia Titan XP GPUs. 
Particularly, the DeDA operation in Eq.~\ref{eq:deda} was implemented in C++ with version 11.1 of CUDA. 
The Adam algorithm \cite{kingma2014adam} with an initial learning rate of $1e-4$ and a multi-step learning rate scheduler with rate halving at 50\%, 70\%, and 90\% of the total epochs, were used for training. 
A mini-batch size of 32 was used for training, and training was stopped after 50 epochs. 
We used three random seeds to train three models for each fold and the final prediction result was determined by ensembing of the logits from three models, followed by a Sigmoid function to produce probability. 
A sensitivity analysis was performed and found out that three random seeds were well in terms of the computational cost versus the performance gain as a balance of all performance metrics.

Data augmentation was performed to enrich the training dataset and improve model generalizability.
Specifically, random flipping, random affine transformation and random Gaussian blurring were used to augment our data.
For augmentation in the training set, lesions were moved to align their center of mass to the geometric center of the image patch. 
Flipping was performed on an orthogonal direction randomly chosen from the axial, coronal, or sagittal direction. 
Affine transformations were performed with a random scale ranging from 0.95 to 1.05 and a random rotation degree between -5 and 5°. 
The final transformed patch was obtained after a trilinear interpolation. 
The blurring was performed using a random-sized Gaussian filter where the kernel radius was determined by $4\sigma + 0.5$. 
The voxel size of our QSM image was $0.75 \times 0.75 \times 3$, thus for the coronal and sagittal direction, we randomly sampled $\sigma \sim \mathcal{N}(0.1,0.95)$, and for the axial direction we randomly sampled $\sigma \sim \mathcal{N}(0.03,0.3)$ for optimal performance.

\subsubsection{More Results forVariants of ResNet and Transformer}

\begin{table}[!t]

\caption{ Performance comparison for transformer networks and deeper convolutional networks.
}

\label{tab:trans}
%\vskip 0.15in
\begin{center}
\resizebox{1\columnwidth}{!}{
\begin{tabular}{lcccccccccc}
\hline
\hline
Backbone	&Pre-Conv	&Pre-Trained	&accu	&f1	&sens	&spec	&ppv	&ROC AUC	&pROC AUC	&PR AUC	\\
\hline											
NesT-Base	&	&	&0.870	&0.265	&0.554	&0.884	&0.174	&0.836	&0.246	&0.194	\\
NesT-Base	&\checkmark	&	&0.963	&0.560	&0.554	&0.981	&0.566	&0.923	&0.619	&0.518	\\
NesT-Base	&\checkmark	&\checkmark	&0.933	&0.446	&0.638	&0.946	&0.342	&0.896	&0.519	&0.406	\\
NesT-Base	&	&\checkmark	&0.869	&0.261	&0.542	&0.884	&0.172	&0.841	&0.253	&0.198	\\
NesT-Small	&	&	&0.912	&0.350	&0.559	&0.927	&0.255	&0.859	&0.359	&0.289	\\
NesT-Small	&	&	&0.959	&0.547	&0.576	&0.976	&0.520	&0.925	&0.617	&0.513	\\
NesT-Small	&	&\checkmark	&0.891	&0.316	&0.593	&0.904	&0.215	&0.843	&0.325	&0.266	\\
NesT-Small	&	&\checkmark	&0.863	&0.251	&0.542	&0.877	&0.164	&0.861	&0.248	&0.187	\\
NesT-Tiny	&	&	&0.862	&0.256	&0.559	&0.875	&0.166	&0.828	&0.237	&0.190	\\
NesT-Tiny	&\checkmark	&	&0.965	&0.565	&0.537	&0.984	&0.597	&0.920	&0.611	&0.511	\\
NesT-Tiny	&\checkmark	&\checkmark	&0.946	&0.447	&0.514	&0.965	&0.396	&0.886	&0.494	&0.391	\\
NesT-Tiny	&	&\checkmark	&0.860	&0.256	&0.565	&0.873	&0.165	&0.827	&0.239	&0.167	\\
\hline											
ViT-Base	&	&	&0.896	&0.295	&0.514	&0.913	&0.207	&0.797	&0.342	&0.264	\\
ViT-Large	&	&	&0.959	&0.545	&0.576	&0.976	&0.518	&0.915	&0.598	&0.497	\\
ViT-Huge	&	&	&0.955	&0.517	&0.565	&0.972	&0.476	&0.909	&0.575	&0.461	\\
\hline											
ResNet-18	&	&	&0.969	&0.641	&0.650	&0.983	&0.632	&0.941	&0.712	&0.612	\\
ResNet-34	&	&	&0.966	&0.634	&0.684	&0.979	&0.590	&0.945	&0.695	&0.587	\\
ResNet-50	&	&	&0.966	&0.621	&0.661	&0.979	&0.585	&0.940	&0.700	&0.584	\\
ResNet-101	&	&	&0.970	&0.639	&0.644	&0.983	&0.633	&0.944	&0.704	&0.597	\\

\hline
\hline
\end{tabular}
}
\end{center}

\end{table}

We included more results on variants of residual and transformer networks.
Two popular transformer networks ViT \cite{dosovitskiyimage} and Swin \cite{liu2021swin} were used to demonstrate the performance.
Particularly we found out that the original Swin transformer performed poorly on the rim dataset, and thus we adopted Nested Transformer (NesT) \cite{zhang2022nested} that was adapted and improved from Swin.
In addition, we further conducted an ablation study on whether to adopt a pre-convolutional (pre-conv) layer (comprises with a $3\times 3 \times 3$ convolution, batch normalization and ReLU activation) or load pre-trained weights for NesT.

As we can see from Table~\ref{tab:trans}, NesT-* with pre-conv layer outperformed their counterparts without such layer, and loading pre-trained weights for all NesT based networks were not useful for the rim+ lesion identification task. 
This can be attributed to the fact that network weights pre-trained on natural images can not be adapted trivially to the new modality adopted in this task.
Considering the difficulty of getting such labeled datasets, it is almost impossible to have domain-specific pre-trained network for this clinical application.
However, interestingly, the best performing model among NesT-* was achieved by NesT-Tiny, which is consistent with results from residual networks \cite{he2016deep} where ResNet-18 achieved the best overall performance (in terms of $F_1$, pROC AUC and PR AUC).
% \newpage

% ---- Bibliography ----

\newpage

\bibliographystyle{splncs04}
\bibliography{mybibliography}

\end{document}